\documentclass{ws-procs9x6}
\usepackage{epsfig}

\def\gsim{\;\rlap{\lower 2.5pt
\hbox{$\sim$}}\raise 1.5pt\hbox{$>$}\;}
\def\lsim{\;\rlap{\lower 2.5pt
   \hbox{$\sim$}}\raise 1.5pt\hbox{$<$}\;}

\begin{document}

%%%%%%%%%%%%%%%%%%%%%%%%%%%%%%%%%%%%%%%%%%%%%%%%%%%%%%%%%%%%%% 
% title, author(s) and address(es) put here:                 %
%%%%%%%%%%%%%%%%%%%%%%%%%%%%%%%%%%%%%%%%%%%%%%%%%%%%%%%%%%%%%% 

\title{The Mass Budget of Merging Quasars}

\author{Kristen Menou}
\address{Institut d'Astrophysique de Paris, 98bis Boulevard Arago, 
75014 Paris, France, menou@iap.fr}  

\author{Zolt\'an Haiman}
\address{Department of Astronomy, Columbia University, 
550 West 120th Street, New York, NY 10027, USA, 
zoltan@astro.columbia.edu}  

%%%%%%%%%%%%%%%%%%%%%%%%%%%%%%%%%%%%%%%%%%%%%%%%%%%%%%%%%%%%%%
% You may repeat \author \address as often as necessary      %
%%%%%%%%%%%%%%%%%%%%%%%%%%%%%%%%%%%%%%%%%%%%%%%%%%%%%%%%%%%%%%

\maketitle

\abstracts{ Two spectacular results emerging from recent studies of
nearby dead quasars and distant active quasars are (i) the existence
of tight relations between the masses of black holes (BHs) and the
properties of their host galaxies (spheroid luminosity or velocity
dispersion, galaxy mass), and (ii) a consistency between the local
mass density in BHs and that expected by summing up the light received
from distant active quasars. These results are partly shaped by
successive galactic mergers and BH coalescences, since mergers
redistribute the population of BHs in galaxies and BH binary
coalescences reduce the mass density in BHs through losses to
gravitational waves.  Here, we isolate and quantify these effects by
following the cosmological merger history of a population of massive
BHs representing the quasar population between $z = 3$ and $z =0$. Our
results suggest that the relation between BH mass and host galaxy
properties, and inferences on the global efficiency of BH accretion
during active quasar phases, could be influenced by the cumulative
effect of repeated mergers.}

%%%%%%%%%%%%%%%%%%%%%%%%%%%%%%%%%%%%%%%%%%%%%%%%%%%%%%%%%%%%%
% The main text of your paper                               %
%%%%%%%%%%%%%%%%%%%%%%%%%%%%%%%%%%%%%%%%%%%%%%%%%%%%%%%%%%%%%

\section{Introduction}

In the past five years, our knowledge of dead and active
quasars\footnote{In what follows, we refer to actively-accreting
massive BHs as ``active quasars'' and to their nearby
inefficiently-accreting counterparts as ``dead quasars.''}  has
improved considerably. For the most part, this is the result of
detailed dynamical studies of nearby galactic nuclei (for dead
quasars) and of several ambitious cosmological surveys (for active
quasars). The progress is so significant that it is now possible to
relate, in a satisfactory manner, the amount of light received from
distant active quasars to the amount of mass locked into nearby dead
quasars.

It has long been suspected that nearby galactic nuclei should host
massive BHs (see Kormendy \& Richstone 1995 for a review) but it is
only recently that the evidence from detailed dynamical studies has
become compelling. BHs appear to be present in the nuclei of almost
all nearby luminous galaxies (Magorrian et al. 1998), and their masses
correlate well with the velocity dispersion of the host galaxy's
spheroidal component (Ferrarese \& Merritt 2000; Gebhardt et al. 2000;
Tremaine et al. 2002; see also Haering \& Rix 2004) and the galaxy's
total mass (Ferrarese 2002). A number of scenarios have been put
forward to explain the origin of these relations between BH mass and
host galaxy properties. They invoke a variety of physical processes,
from radiative and/or mechanical feedback acting during BH formation
(Silk \& Rees 1998; Haehnelt, Natarajan \& Rees 1998; Adams, Graff \&
Richstone 2001; King 2003; Wyithe \& Loeb 2003b; Di Matteo et
al. 2003) to dynamical processes operating in the BH environment
(Ostriker 2000; Zhao, Haehnelt \& Rees 2002; Merritt \& Poon 2004;
Miralda-Escude \& Kollmeier 2004; Sellwood \& Moore 1999; Shen \&
Sellwood 2003).

Well before the case for a local population of massive BHs was made,
it had been proposed that dead quasars, the remnants of past active
quasar phases, should be present in today's galactic nuclei, and that
the amount of mass locked into these BHs should be directly related to
the amount of light released by distant active quasars (Lynden-Bell
1969; Soltan 1982; Rees 1990). By combining recent characterizations
of the local population of massive BHs with high-quality data on
nearby galaxies and distant active quasars from modern cosmological
surveys (2dF: Boyle et al. 2000; SDSS: Stoughton et al. 2002), it is
now possible to confirm the link between active and dead quasars with
surprising accuracy.

Yu \& Tremaine (2002) find a mass density $\rho_{\rm BH} \simeq 2.5
\times 10^5$~M$_\odot$~Mpc$^{-3}$ for the local population of massive
BHs (see also Aller \& Richstone 2002). On the other hand, assuming a
mass-to-light conversion efficiency $\epsilon =0.1$ for BH accretion,
they also infer a mass density $\rho_{\rm BH} \simeq 2.1 \times
10^5$~M$_\odot$~Mpc$^{-3}$ from the integrated light of
optically-bright quasars (see also Chokshi \& Turner 1992). The
consistency between these two values is remarkable and it provides an
effective measure of the BH accretion efficiency during
(optically-bright) active quasar phases. It appears, however, that
some of the light from active quasars may have been missed by optical
surveys (probably because of dust obscuration), since several authors
have inferred values of $\rho_{\rm BH}$ from X-ray data that exceed
substantially the corresponding value for optical data (Fabian \&
Iwasawa 1999; Barger et al. 2001; Elvis, Risaliti \& Zamorani
2002). Requiring consistency with the local value of $\rho_{\rm BH}$
then points toward a larger efficiency for BH accretion, $\epsilon
\gsim 0.15$--$0.2$.  As the radiative efficiency can increase for gas
accreting onto spinning BHs, this may indicate that massive BHs
associated with luminous quasars are spinning rapidly.

These impressive developments have focused on the ``brighter side'' of
quasar evolution, the one that is most easily accessible to
astronomers through various electromagnetic signatures of accretion
onto massive BHs. Our main interest here is the ``darker side'' of
quasar evolution, which is (currently) not easily accessible to
astronomers, but could in principle be equally important in shaping
the properties of the quasar population. It is a generic prediction of
hierarchical CDM cosmologies that galaxies merge to grow larger and
more massive with cosmic times and, as a result, the massive BHs that
they harbor are expected to coalesce.  BH coalescences should be
observable in the future through the detection of gravitational waves
accompanying these events.  Studying this darker side of quasar
evolution is in fact one of the main motivations behind efforts to
build the Laser Interferometer Space Antenna (LISA), and several
studies have already emphasized how valuable the information provided
by LISA on the population of distant quasars would be (Haehnelt 1994;
2003; Menou, Haiman \& Narayanan 2001; Hughes 2002; Menou 2003; Sesana
et al. 2004; Islam, Taylor and Silk 2004; Wyithe \& Loeb 2003a).

Here, we wish to investigate the consequences that galactic mergers
and BH coalescences may have on the cosmological evolution of
quasars. Specifically, we focus on two important ways in which mergers
can potentially affect recent results on the populations of dead and
active quasars.  First, in each BH binary coalescence, a finite amount
of energy is lost to gravitational waves. This translates into an
effective mass loss for the remnant BH, which ends up with a mass
smaller than the initial mass of its two progenitors. Although this
mass loss can be as small as $\sim 6 \%$ of the lowest mass
progenitor, according to general relativity (see \S2.3), it can also
be significantly larger.  Most importantly in a cosmological context,
the effect is cumulative, since each individual BH experiences mergers
repeatedly over time. Second, successive galactic mergers effectively
redistribute the population of massive BHs in galaxies, thus changing
its overall properties. For example, it is unclear whether the
characteristics of a high-redshift quasar population initially similar
to those of nearby dead quasars would be conserved or modified by
successive cosmological mergers (see, e.g., Ciotti \& van Albada 2001;
Haehnelt 2003; Koushiappas, Bullock \& Dekel 2004).

In the present work, we isolate and quantify these two effects by
following the cosmological merger history of a plausible population of
quasars. We describe our methodology in detail in \S2 and discuss our
results in \S3.

\section{Models}

\subsection{General Characteristics}

To follow the cosmological merger history of a population of quasars,
we must first describe how the galaxies hosting them evolve with
cosmic times. We do this by using Monte--Carlo simulations of ``merger
trees'' that describe the merger history of dark matter halos (and
associated galaxies) in the standard $\Lambda$CDM cosmology ($\Omega_0
=0.3$, $\Omega_b=0.04$, $\Omega_\Lambda=0.7$, $h_{100}=0.65$). The
specific merger tree we are using, and a number of additional model
assumptions, are described in detail in Menou et al. (2001). The range
of halo masses considered (e.g. $\sim 10^9$--$10^{13}$~M$_\odot$ at
$z=0$, for a fixed comoving volume of $\sim 1.7 \times 10^4$~Mpc$^3$)
guarantees that one-to-one associations between dark matter halos and
galaxies are reasonably accurate (Menou et al. 2001).  One of the main
model assumptions is that only galaxies with a virial temperature in
excess of $10^4$~K are able to host a BH because they are the only
ones in which baryon cooling is efficient enough to allow BH formation
(see, e.g., Loeb \& Barkana 2001 and Haiman \& Quataert 2004 for
reviews). By construction, the merger tree describes only
``interesting'' halos with virial masses in excess of a
temperature-equivalent of $10^4$~K.

The complex and uncertain physics of baryon cooling and galaxy
formation is not described by our models. As a result, no attempt is
made to separate from the rest a bulge-less galactic population, which
may or may not be able to harbor massive BHs according to recent
studies (Gebhardt et al. 2001; Merritt et al. 2001, but see Salucci et
al. 2000). Accounting for several different galactic types in our
simulations would further complicate their interpretation and we have
chosen, for simplicity, to assume that every halo described by the
tree is a possible host for a massive BH.

Currently, observational constraints on the presence of massive BHs in
low-mass galaxies are very scarce and various arguments have been put
forward to suggest that these galaxies may not be able to form, or
perhaps retain, massive BHs (e.g. Haiman, Madau \& Loeb 1999;
Ferrarese 2002; Haiman, Quataert \& Bower 2004; Haehnelt et al. 1998;
Silk \& Rees 1998; Favata, Hughes \& Holz 2004; Merritt et al. 2004;
Madau \& Quataert 2004; Bromley, Somerville \& Fabian 2004; but see
also Barth et al. 2004). In order to maximize the cumulative effects
of successive BH mergers, we assume by default that a massive BH is
present in all the galaxies described by the merger tree. For the sake
of generality, however, we will also investigate alternative scenarios
in which BHs preferentially populate massive galaxies, at least
initially.

A realistic model for the cosmological evolution of a population of
quasars should describe simultaneously the action of mergers and
accretion on the population of massive BHs. Such models exist already
(e.g. Kauffmann \& Haehnelt 2000; Volonteri, Haardt \& Madau 2003) and
their focus generally is on reproducing a number of observational
constraints available for the bright side of quasar evolution.
However, these studies have not isolated the contribution due to
mergers and have not included gravitational wave losses when computing
the evolution of BH masses.  Our goal here is not to construct a
realistic evolution scenario for quasars, but precisely to isolate and
quantify the effects that mergers and coalescences may have on the
quasar population.

Recent quasar evolutionary studies indicate that massive BHs acquired
most of their accreted mass over a wide range of redshifts,
approximately $90 \%$ of it from $z \simeq 3$ to $z \simeq 0$ (see,
e.g., Fig.~1 of Yu \& Tremaine 2002, or Fig.~8 of Marconi et
al. 2004). Consequently, both mergers and accretion will act to shape
the main properties of the quasar population over this range of
redshifts.  In our models, we take no account of accretion, and
instead assume that the population of massive BHs present at $z=3$
evolves from $z=3$ to $z=0$ only through a hierarchy of mergers. We
emphasize that this assumption is not intended to yield a realistic
description of the observed quasar population.  Rather, the motivation
behind this assumption is to provide a robust clarification of the
role of mergers alone.

Making this assumption still leaves the detailed characteristics of
the quasar population in place at $z=3$ completely open, in the sense
that the BH mass can be distributed in host galaxies in many different
ways. The results of Shields et al. (2003) indicate, however, that the
properties of massive BHs associated with luminous quasars at $z
\simeq 3$ are consistent with the properties of dead quasars studied
locally. Based on this result, we have chosen to investigate the
effects of mergers on a population of quasars with characteristics
similar to those of local dead quasars. In the next subsection, we
describe in further detail how BH masses were chosen with respect to
the mass of host galaxies in our models.

\subsection{Black Hole Masses}

We consider two different, albeit related models for the initial
distribution of BH masses in galaxies at $z=3$. Our main motivation
for exploring two mass models is the possibility that our results (at
$z=0$) are sensitive to the exact characteristics of the BH population
assumed to be present at $z=3$.  In the first class of models (``T
models''), BH masses are chosen according to the relation established
by Tremaine et al. (2002),
\begin{equation}
M_{\rm bh}=\left( 1.35 \pm 0.2 \right) \times 10^8 M_\odot \left(
\frac{\sigma_e}{200~{\rm km~s^{-1}}} \right)^{4.02 \pm 0.32},
\label{eq:one}
\end{equation}
where $\sigma_e$ is the stellar velocity dispersion of the spheroidal
component at the half-light (effective) radius. It is related to
$\sigma_{\rm dm}$, the halo velocity dispersion, via the relation
$\sigma_e = \sigma_{\rm dm}/\sqrt{3/2}$, which is derived from the
Jeans equation for isotropic, spherical systems, assuming an
isothermal density profile ($\rho \propto r^{-2}$) for the dark matter
and a typical DeVaucouleurs density profile ($\rho \propto r^{-3}$)
for the stellar spheroidal component.  The halo velocity dispersion is
obtained from the virial theorem and the assumption that halos have a
universal density,
\begin{equation}
\sigma_{\rm dm} = 17 \left( \frac{M_{\rm halo}}{10^8 M_\odot h}
\right)^{1/3} \left( \frac{\delta_c}{200} \right)^{1/6} \left(
\frac{1+z}{10} \right)^{1/2},
\end{equation}
where $\delta_c (z)$ is the cosmological density contrast at
collapse. This is the exact same prescription as the one adopted in
Menou (2003).

In the second class of models (``FWL models''), BH masses are chosen
according to the relation established locally by Ferrarese (2002; with
a different $\sigma_e$--$\sigma_{\rm dm}$ relation than above) and
extended to higher redshifts by Wyithe \& Loeb (2004),
\begin{equation}
M_{\rm bh}= 10^9 M_\odot \left( \frac{M_{\rm halo}}{1.5 \times 10^{12}
M_\odot}\right)^{5/3} \left( \frac{1+z}{7} \right)^{5/2}.
\end{equation}
A similar prescription (with a somewhat smaller normalization) has
been adopted by Haiman, Quataert \& Bower (2004) for their predictions
on high-redshift radio-loud quasars.

The two prescriptions (T and FWL) have a similar dependence on
redshift ($(1+z)^2$ vs. $(1+z)^{5/2}$) but different normalizations
and power-law scalings with $M_{\rm halo}$ ($\sim 4/3$ for T vs. $5/3$
for FWL). Another difference is the initial presence of scatter in the
T relation (see equation~[\ref{eq:one}]), which is included in our T
models but is absent from FWL models.

\subsection{Gravitational Wave Losses}

When two galaxies hosting massive BHs merge, several processes act
successively to bring the BHs closer to the galactic remnant's center,
have them form a bound binary and ultimately make them coalesce
(Begelman, Blandford \& Rees 1980). The first such process is
dynamical friction, which is thought to be rather efficient initially
(but see discussion in \S2.4). When the separation between the two BHs
reduces to parsec scales typically, other mechanisms must be invoked,
however (e.g. repeated stellar ejections, gaseous interaction). These
mechanisms have been investigated in detail (see, e.g., Milosavljevic
\& Merritt 2003; Blaes, Lee \& Socrates 2002; Gould \& Rix 2000;
Armitage \& Natarajan 2002 for recent results) but the efficiency with
which they act to shrink the orbit of a massive BH binary is still
much debated. Ultimately, at sub-parsec separations, emission of
gravitational waves takes over as the leading mechanism for the loss
of energy and angular momentum from the binary, until the two BHs
coalesce.

As we have already emphasized, our main interest here is in isolating
the effects of mergers on a quasar population. For this reason, we
will neglect complexities related to the various processes we have
just mentioned and assume by default in our models that two BHs
coalesce efficiently right after their host galaxies merge. This
assumption simplifies the models greatly and it is consistent with our
effort to quantify the effects that many successive mergers may have
on a quasar population.

During their final shrinking phases, massive BH binaries will
therefore lose energy in the form of gravitational waves, but the
total amount of mass-energy lost after coalescence is complete is not
well known.  A first source of uncertainty arises from general
relativistic calculations, because of the non-linear character of the
strong-field interaction between the two BHs.  A second source of
uncertainty comes from the necessity of knowing the masses and the
spins of the two BHs involved, as well as the orbital geometry, to
make accurate predictions. While the masses of BHs in nearby dead
quasars are known with some accuracy, masses of BHs in more distant
quasars are not so well known and essentially no information is
available on BH spins. In the context of quasar evolution, where both
mergers and accretion contribute to the coupled mass and spin
evolution of BHs (in a way which has yet to be elucidated; see, e.g.,
Gammie, Shapiro \& McKinney 2004 and Hughes \& Blandford 2003 for
recent discussions), it is not possible to provide an accurate
description of the spin evolution of quasar BHs.

The coalescence of two BHs is often decomposed in three successive
phases (see, e.g., Hughes 2002 and references therein): the slow
inspiral phase (when the two BHs spiral in quasi-adiabatically), the
dynamical plunge phase (when the two BHs plunge towards each other and
their horizons merge) and the final ringdown phase (when the merger
remnant relaxes to a stationary Kerr BH solution). Emission of
gravitational waves during the plunge phase is not well understood
because it is not well approximated by perturbative methods. Small
losses during this phase are generally expected because of its short
duration. Significant progress with numerical simulations has been
made in recent years but the problem has not been addressed yet in
full generality (see Baumgarte \& Shapiro 2003 for a review).  The
emission of gravitational waves during the ringdown phase is also
subject to uncertainties, since its initial condition is the unknown
outcome of the plunge phase, but it is also expected to contribute a
relatively small amount to the total energy lost during coalescence
(Khanna et al. 1999).

Emission of gravitational waves during the inspiral phase is
comparatively much better understood. Losses are calculated by
identifying the location of the Innermost Stable Circular Orbit
(ISCO), which corresponds to the minimum separation that the two
massive BHs are able to reach via slow inspiral before plunge.  For
example, for very small mass ratios (approaching the test particle
case), it is well known that the binary loses the equivalent of $\sim
6 \%$ of the rest mass of the less massive BH if the more massive one
is non-rotating, and $\sim 42 \%$ of the rest mass of the less massive
BH if the more massive one is maximally-rotating (from arguments
similar to those yielding the radiative efficiency of gas accretion
onto BHs, see, e.g. Shapiro \& Teukolsky 1983). Losses during the
subsequent plunge and ringdown phases would have to be added to these
figures.

Independently of the details of the merger process, the BH area
theorem of general relativity also puts strict limits on the total
amount of energy lost to gravitational waves when two BHs
coalesce. Although it is likely that these bounds largely overestimate
the actual losses in astrophysical coalescences,\footnote{This can
been shown explicitly for the head-on collision of two Schwarzschild
BHs by comparing losses in direct numerical integrations with the
corresponding upper limit provided by general relativity (see Shapiro
\& Teukolsky 1983).} they are still useful in providing firm upper
limits on the cumulative effects of mergers on a quasar population.

In view of all the above uncertainties, we explore five simple
prescriptions for gravitational wave losses in our models. In a given
evolutionary model for the quasar population, the same prescription is
adopted for all BH binary coalescences.  Let us denote by $M_1$ the
mass of the most massive BH involved and $M_2$ that of the lowest mass
BH ($M_1 =M_2$ never occurs exactly in our models, even if sometimes
$M_1 \simeq M_2$).

\begin{enumerate}
\item In models with suffix ``0,'' losses to gravitational waves are
neglected and the masses $M_1$ and $M_2$ are simply added up during
coalescences. The interest of this model is that it allows us to
isolate the effects of galactic mergers on the quasar population,
independently of gravitational wave losses.\\

\item In models with suffix ``6,'' losses to gravitational waves are
taken to be $6 \%$ of $M_2$. This corresponds to a situation in which
the largest mass BH ($M_1$) is non-rotating (Schwarzschild), the mass
ratio is assumed to be very small ($M_2/M_1 \ll 1$) and losses during
the plunge and ringdown phases are neglected.\\

\item In models with suffix ``42,'' losses to gravitational waves are
taken to be $42 \%$ of $M_2$. This corresponds to a situation in which
the largest mass BH ($M_1$) is maximally rotating (max-Kerr), the mass
ratio is assumed to be very small ($M_2/M_1 \ll 1$) and losses during
the plunge and ringdown phases are neglected.\\

\item In models with suffix ``adS,'' the maximum losses allowed by the
BH area theorem for the coalescence of two Schwarzschild BHs resulting
in a Schwarzschild BH are adopted.  The corresponding mass deficit is
$\Delta M = M_1+M_2 -\sqrt{M_1^2 + M_2^2}$, with a maximum of $29 \%$
when $M_1=M_2$ (e.g., Shapiro \& Teukolsky 1983).\\

\item In models with suffix ``adK,'' the maximum losses allowed by the
BH area theorem for the coalescence of two counter-rotating max-Kerr
BHs resulting in a Schwarzschild BH are adopted.  The corresponding
mass deficit is $\Delta M = M_1+M_2 -\sqrt{(M_1^2 + M_2^2)/2}$, with a
maximum of $50 \%$ when $M_1=M_2$ (e.g., Shapiro \& Teukolsky 1983).
 
\end{enumerate}

\subsection{Inefficient Dynamical Friction}

By accounting for the gradual tidal evaporation of the stellar cluster
initially bound to a massive BH which experiences dynamical friction
in a realistic galaxy model, Yu (2002) has argued that binary BHs with
mass ratios $M_2/M_1 < 10^{-3}$ are unable to form: the dynamical
friction time for the smallest mass BH ($M_2$) then exceeds a Hubble
time. Although this argument has been developed for local galaxies,
similar conclusions may hold for galaxies at higher redshifts. It
would imply that massive BH binaries with large mass ratios are unable
to merge and it could thus potentially influence our study of the
cumulative effects of mergers on a quasar population.

To test the influence of inefficient dynamical friction, we have also
constructed models in which massive BH binaries are not allowed to
coalesce unless their mass ratio is large enough.
\begin{enumerate}
\item In models with suffix ``q3,'' following Yu (2002), we assume
that BH binaries coalesce only if $M_2/M_1 > 10^{-3}$. In galactic
mergers involving BHs which do not satisfy the above constraint, the
lowest mass BH ($M_2$) is simply ignored from the subsequent
cosmological evolution. Losses to gravitational waves are also ignored
for all BH coalescences, so that the effect of inefficient dynamical
friction can be isolated.\\

\item In models with suffix ``q2,'' we assume that dynamical friction
is even less efficient than above and allow BH binaries to coalesce
only if $M_2/M_1 > 10^{-2}$.
\end{enumerate}

\section{Results}

\subsection{Forced Models}

\begin{figure}[t]
\begin{center}
\hspace{-1cm}
\begin{minipage}[t]{1.00\hsize}
\begin{displaymath}
\psfig{figure=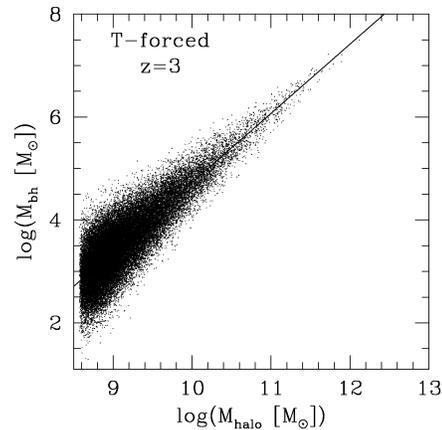,height=2.3in}
\end{displaymath}
\end{minipage}
\end{center}
\caption{\label{fig:one} Initial distribution of black hole masses
($M_{\rm bh}$) and halo masses ($M_{\rm halo}$) at $z=3$ for a
population of quasars following a relation adopted from Tremaine et
al. (2002; T-forced at $z=3$ in the notation of Table~1). The solid
line is a least-square fit to the distribution (see Table~1 for
parameters).}
\end{figure}

Figure~\ref{fig:one} shows the initial distribution of $M_{\rm
bh}$--$M_{\rm halo}$ in a T-model at $z=3$. Halo masses range from
$\sim 3 \times 10^8$~M$_\odot$ to $\sim 10^{12}$~M$_\odot$ in the
merger tree at this redshift and the corresponding range of BH masses
is $\sim 10^2$~M$_\odot$ to $3 \times 10^7$~M$_\odot$, with
significant scatter for a given halo mass (from Eqs.~[1] and [2]).
Given the fixed comoving volume of $\sim 1.7 \times 10^4$~Mpc$^3$
described by the merger tree, it is possible to sum up all the BH
masses in the model and deduce a mass density in BHs at $z=3$:
$\rho_{\rm BH} \simeq 7.3 \times 10^4$~M$_\odot$~Mpc$^{-3}$. This
value is about a factor four smaller than the value measured locally,
even though we have used an observationally determined relation
(Eq.~[1]) to populate our halos with BHs. The origin of this
discrepancy lies, in fact, in the limited number of simulated halos
described by the merger tree that we are using.\footnote{The tree size
is limited, in practice, by the numerous small mass halos close to the
mass threshold for efficient baryon cooling at $T_{\rm vir} =
10^4$~K.}  As can be seen from Fig.~\ref{fig:one}, massive halos are
scarce (e.g. only one halo in excess of $10^{12}$~M$_\odot$ at
$z=3$). Since the BH mass function is dominated by large mass BHs, the
statistical scarcity of massive halos in the tree results in a
somewhat underestimated value of $\rho_{\rm BH}$. This situation does
not limit the predictive power of our models, however, as long as we
are careful enough to characterize how our results depend on halo
masses. A similar exercise for a population of quasars following the
FWL relation (Eq.~[3]), instead of the T relation, at $z=3$ results in
a value of $\rho_{\rm BH} \simeq 1.4 \times
10^5$~M$_\odot$~Mpc$^{-3}$, only a factor of $\sim 2$ times smaller
than the measured value.

In order to quantify the results of our simulations for the $M_{\rm
bh}$--$M_{\rm halo}$ relation, we have found it useful to perform
least-square fits to distributions such as the one shown in
Fig.~\ref{fig:one}. Assuming a dependence of the form
\begin{equation}
\log (M_{\rm
bh}) = n \times \log (M_{\rm halo}) + \alpha, 
\end{equation}
where both $M_{\rm bh}$ and $M_{\rm halo}$ are expressed in solar
units, the least-square algorithm provides us with best fit values for
$n$, $\alpha$ and the residual scatter, $\sigma$, around the best fit
line.  Often, our results show significant dependence on $M_{\rm
halo}$, so we also perform least square fits restricted to masses
$M_{\rm halo} > 10^{11}$~M$_\odot$. The parameters resulting from
these restricted fits are noted $n_{11}$, $\alpha_{11}$ and
$\sigma_{11}$. Note that our goal in performing these fits is not to
provide accurate descriptions of the $M_{\rm bh}$--$M_{\rm halo}$
distribution, but simply to provide quantitative means to compare
results from different models.

\begin{table}[t]
\ttbl{30pc}{MODEL PROPERTIES}
{\begin{tabular}{ccrrrrrr}\\
\multicolumn{8}{c}{(Mass Density and LSQ--Fit Parameters)} \\[6pt]\hline
\\
Model & $\rho_{\rm BH}$ & $n_{11}$ & $\alpha_{11}$ & $\sigma_{11}$& $n$ & $\alpha$ & $\sigma$\\
&(M$_\odot$ Mpc$^{-3}$)& & & & & &\\
\\
\hline
\\
T--forced at $z=3$ & $7.3 \times 10^4$& $1.36$&$-8.90$& $0.16$&  $1.35$&$-8.73$& $0.37$\\
FWL--forced at $z=3$ & $1.4 \times 10^5$&$1.67$&$-11.90$&$0.00$& $1.67$&$-11.90$&$0.00$\\
Trare--forced at $z=3$ & $6.0 \times 10^4$&$1.34$&$-8.71$&$0.14$& $1.34$&$-8.68$&$0.28$\\
\\
\hline
\\
T0 &$7.3 \times 10^4$&$1.19$ &$-7.47$& $0.13$ &$1.38$&$-9.62$&$0.37$ \\
FWL0 &$1.4 \times 10^5$& $1.46$& $-10.42$& $0.19$&$1.57$&$-11.70$& $0.26$\\
Trare0 &$6.0 \times 10^4$& $1.30$& $-8.82$& $0.19$&$1.27$&$-8.58$&$0.31$\\
T6 &$6.9 \times 10^4$& $1.18$& $-7.41$& $0.14$&$1.37$&$-9.52$&$0.37$\\
T42 &$5.0 \times 10^4$& $1.14$& $-7.05$& $0.18$&$1.30$&$-8.85$&$0.37$\\
FWL42 &$1.1 \times 10^5$& $1.43$& $-10.16$& $0.25$&$1.50$&$-11.11$&$0.28$\\
Trare42 & $4.4 \times 10^4$&$1.22$&$-7.98$&$0.22$& $1.18$& $-7.65$& $0.32$\\
TadS &$3.4 \times 10^4$& $1.07$& $-6.46$& $0.24$&$1.21$&$-8.03$&$0.38$\\
TadK &$3.8 \times 10^3$&$-0.27$ &$7.31$ & $0.31$ &--&--&--\\
Tq3 &$7.2 \times 10^4$& $1.19$& $-7.44$& $0.13$&$1.38$&$-9.59$&$0.37$\\
Tq2 &$7.0 \times 10^4$& $1.18$& $-7.33$& $0.13$&$1.38$&$-9.59$&$0.37$\\
\\
\hline
\end{tabular}}
\label{tab:one}
\end{table}

Table~\ref{tab:one} lists the properties of most of the models we have
explored in this study. In each case, the value of the BH comoving
mass density, $\rho_{\rm BH}$, and the various fit parameters are
given. The first group of models in Table~\ref{tab:one} (first 3
lines) corresponds to models in which the quasar population was forced
to follow one or the other prescription for BH masses (T-forced or
FWL-forced) at $z=3$. The second group of models (last 11 lines)
corresponds to ``evolutionary'' models in which the quasar population
was forced to follow only initially one or the other prescriptions for
BH masses (at $z=3$). After that, the quasar population is evolved
through successive cosmological mergers according to one of the
prescriptions for gravitational wave losses or BH coalescences defined
in \S2. For this second group of models, properties are only listed in
Table~\ref{tab:one} at $z=0$, after cosmological evolution is
complete.

The first group of ``forced'' models is useful to illustrate a number
of general properties and for comparison with the second group of
``evolutionary'' models. Least square fits to forced models, such as
the one shown in Fig.~\ref{fig:one} (T-forced at $z=3$), recover the
correct value for the slope of the $M_{\rm bh}$--$M_{\rm halo}$
relation ($n$ and $n_{11} \simeq 4/3$ for T-forced models and $\simeq
5/3$ for FWL-forced models; see Table~\ref{tab:one}). They also show
that the scatter in the distribution is dominated by small masses
($\sigma_{11} > \sigma$), as is visually suggested by
Fig.~\ref{fig:one}.

\subsection{Evolutionary Models}

\begin{figure}[t]
\begin{center}
\hspace{-1.7cm}
\begin{minipage}[t]{0.53\hsize}
\begin{displaymath}
\psfig{figure=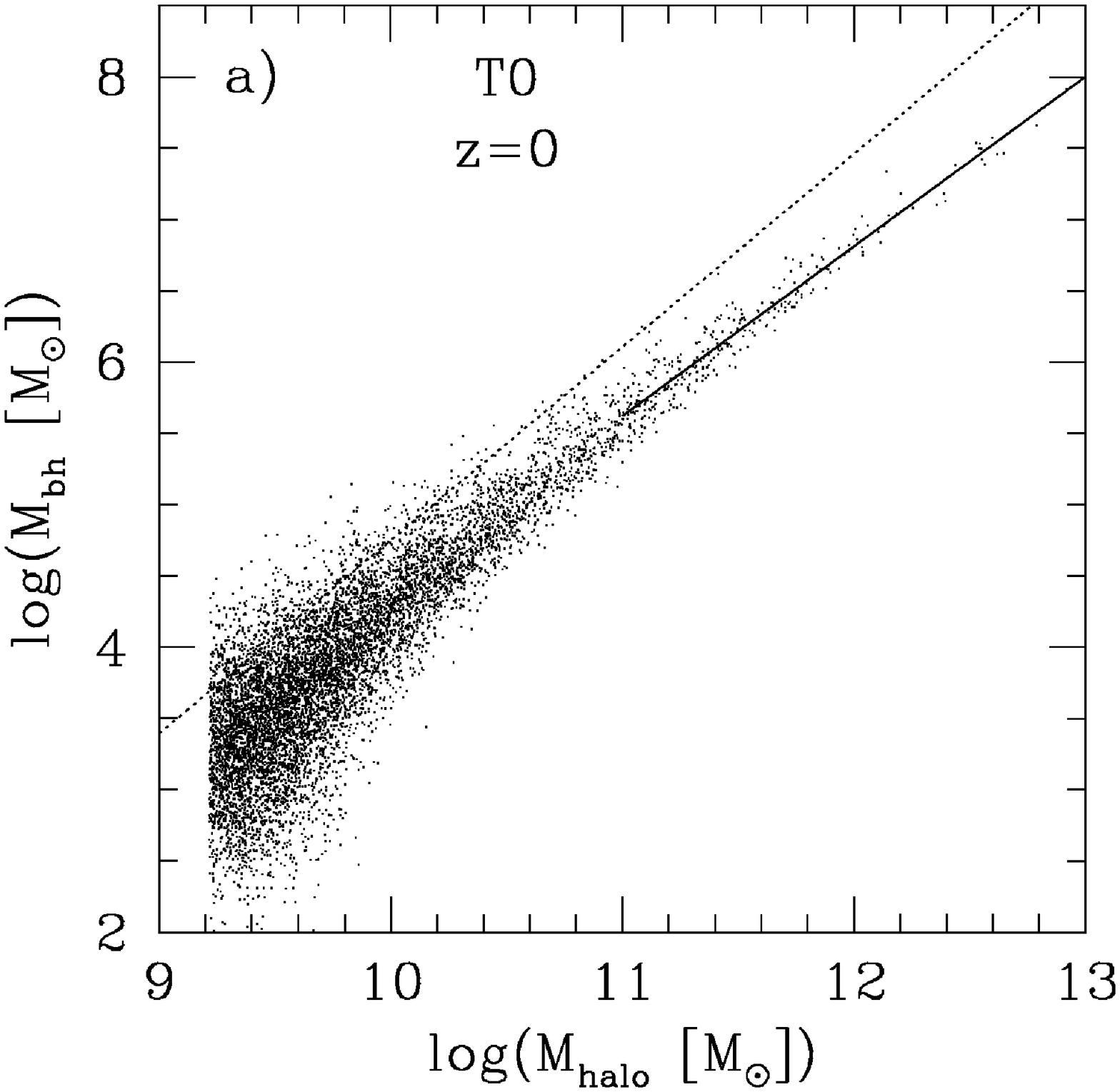,height=2.3in}
\end{displaymath}
\end{minipage}
\begin{minipage}[t]{0.60\hsize}
\begin{displaymath}
\psfig{figure=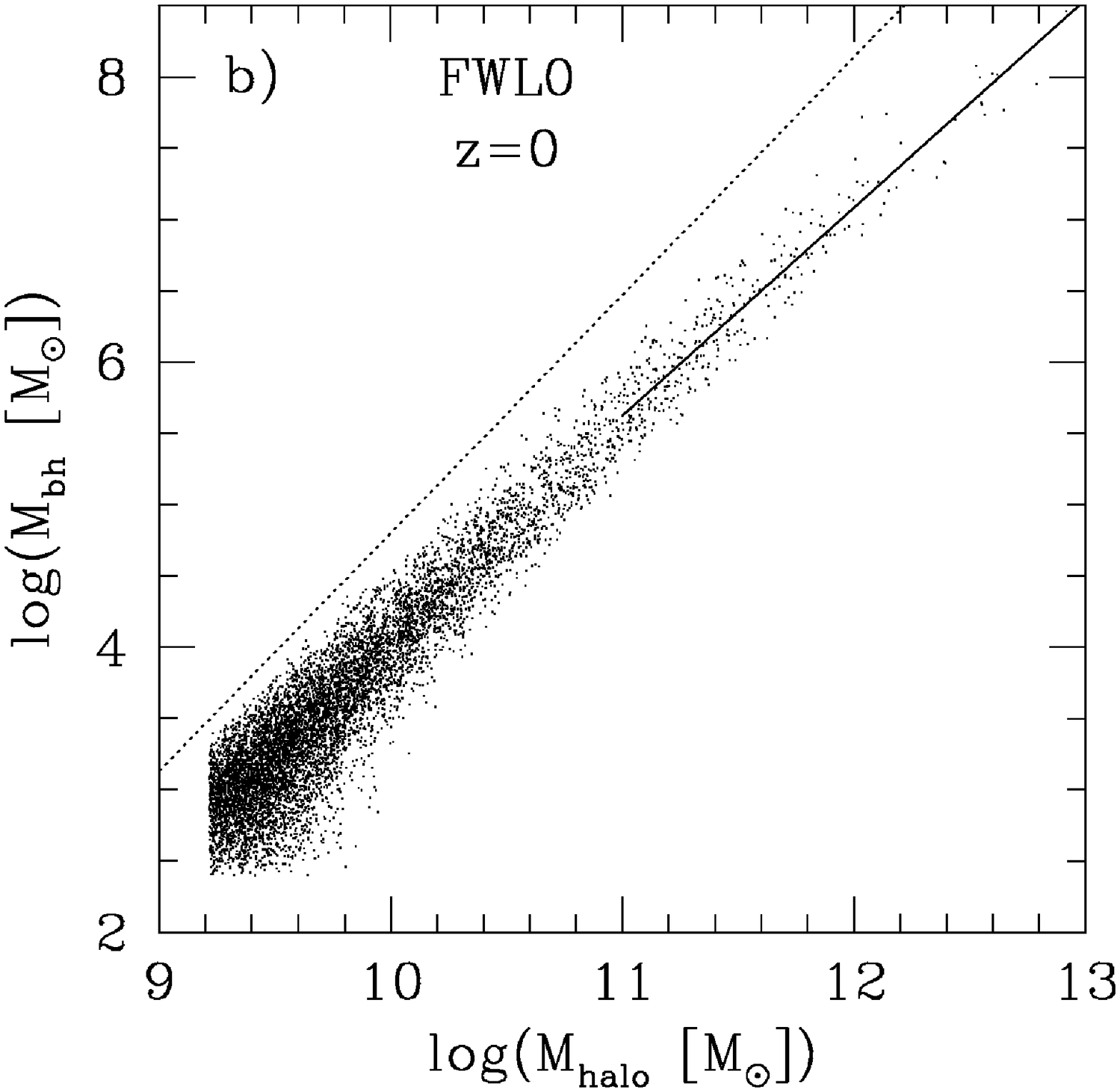,height=2.3in}
\end{displaymath}
\end{minipage}
\end{center}
\caption{\label{fig:two} $M_{\rm bh}$--$M_{\rm halo}$ end
distributions for populations of quasars having experienced a series
of cosmological mergers from $z=3$ to $z=0$. Efficient black hole
coalescence following a galactic merger is assumed and energy losses
to gravitational waves are neglected. Solid lines are least-square
fits to the distributions, restricted to $M_{\rm halo} >
10^{11}~M_\odot$ (see Table~1 for fit parameters). Model T (a)
corresponds to a population initially following a relation adopted
from Tremaine et al. (2002) at $z=3$, while model FWL corresponds to a
population initially following a relation adopted from Ferrarese
(2002) and Wyithe \& Loeb (2004). Dotted lines are least-square fits
obtained for populations which are forced to follow the T or FWL
relations at $z=3$. The cumulative effect of mergers is to flatten the
$M_{\rm bh}$--$M_{\rm halo}$ relation, especially at large masses.}
\end{figure}

Figure~\ref{fig:two} shows the $M_{\rm bh}$--$M_{\rm halo}$ end
distributions which result when quasar populations are forced only
initially to follow either the T-relation (a) or the FWL-relation (b)
at $z=3$ and are subsequently left to evolve via cosmological mergers,
without any mass loss to gravitational waves. In this case, the total
mass locked into BHs is conserved over cosmic times and the value of
$\rho_{\rm BH}$ at $z=0$ is the same as the original value at $z=3$
($\rho_{\rm BH} \simeq 7.3 \times 10^4$~M$_\odot$~Mpc$^{-3}$ for the
T0 model and $\simeq 1.4 \times 10^5$~M$_\odot$~Mpc$^{-3}$ for the
FWL0 model; see Table~\ref{tab:one}).

Still, the quasar population is redistributed among galaxies of
different masses through cosmological mergers and this modifies the
$M_{\rm bh}$--$M_{\rm halo}$ distribution in both models. Fit
parameters for the entire distribution and for the distribution
restricted to $M_{\rm halo} > 10^{11}$~M$_\odot$ are given in
Table~\ref{tab:one} for both models. Solid lines in
Fig.~\ref{fig:two}a and~\ref{fig:two}b, which are the best fit lines
to the restricted distributions, have significantly flatter slopes
than the best fit lines for forced models at $z=3$ (dotted
lines). This shows that the cumulative effect of mergers (without any
contribution from gravitational wave losses) is to flatten the $M_{\rm
bh}$--$M_{\rm halo}$ relation. The effect is clearly stronger at large
masses since the slope difference is much less pronounced if one
considers the entire $M_{\rm bh}$--$M_{\rm halo}$ distributions (see
fit parameters in Table~\ref{tab:one}).  This result arises because
the more massive halos have experienced a larger number of
mergers. Under the assumption that the two BH masses simply add
together in a merger, each merger event will cause the resulting BH
mass to fall below the $M_{\rm bh}$--$M_{\rm halo}$ relation.

The overall normalization of the two end distributions shown in
Fig.~\ref{fig:two}, below the dotted lines, is not very meaningful.
Note that the forced models at $z=3$ (dotted lines) have the exact
same value of $\rho_{\rm BH}$ as the $M_{\rm bh}$--$M_{\rm halo}$
distributions at $z=0$ shown in Fig.~\ref{fig:two}a and~\ref{fig:two}b
even if it is not obviously apparent (compare T0 and FWL0 with
T-forced and FWL-forced at $z=3$ in Table~\ref{tab:one}). What is
significant, however, is the difference in slope of the $M_{\rm
bh}$--$M_{\rm halo}$ relation, especially at large masses, which
results purely from cosmological mergers.

\begin{figure}[t]
\begin{center}
\hspace{-1.7cm}
\begin{minipage}[t]{0.53\hsize}
\begin{displaymath}
\psfig{figure=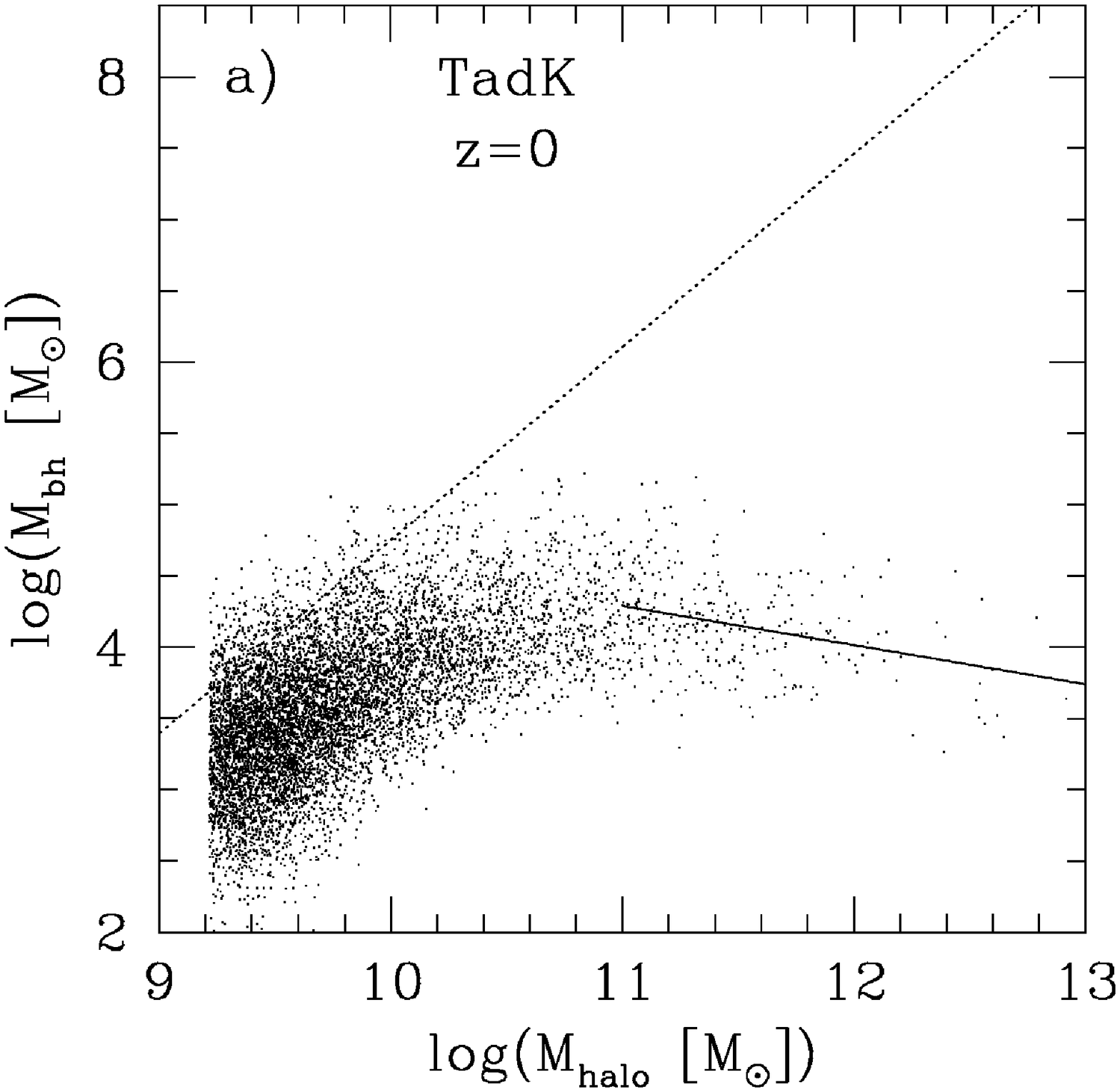,height=2.3in}
\end{displaymath}
\end{minipage}
\begin{minipage}[t]{0.60\hsize}
\begin{displaymath}
\psfig{figure=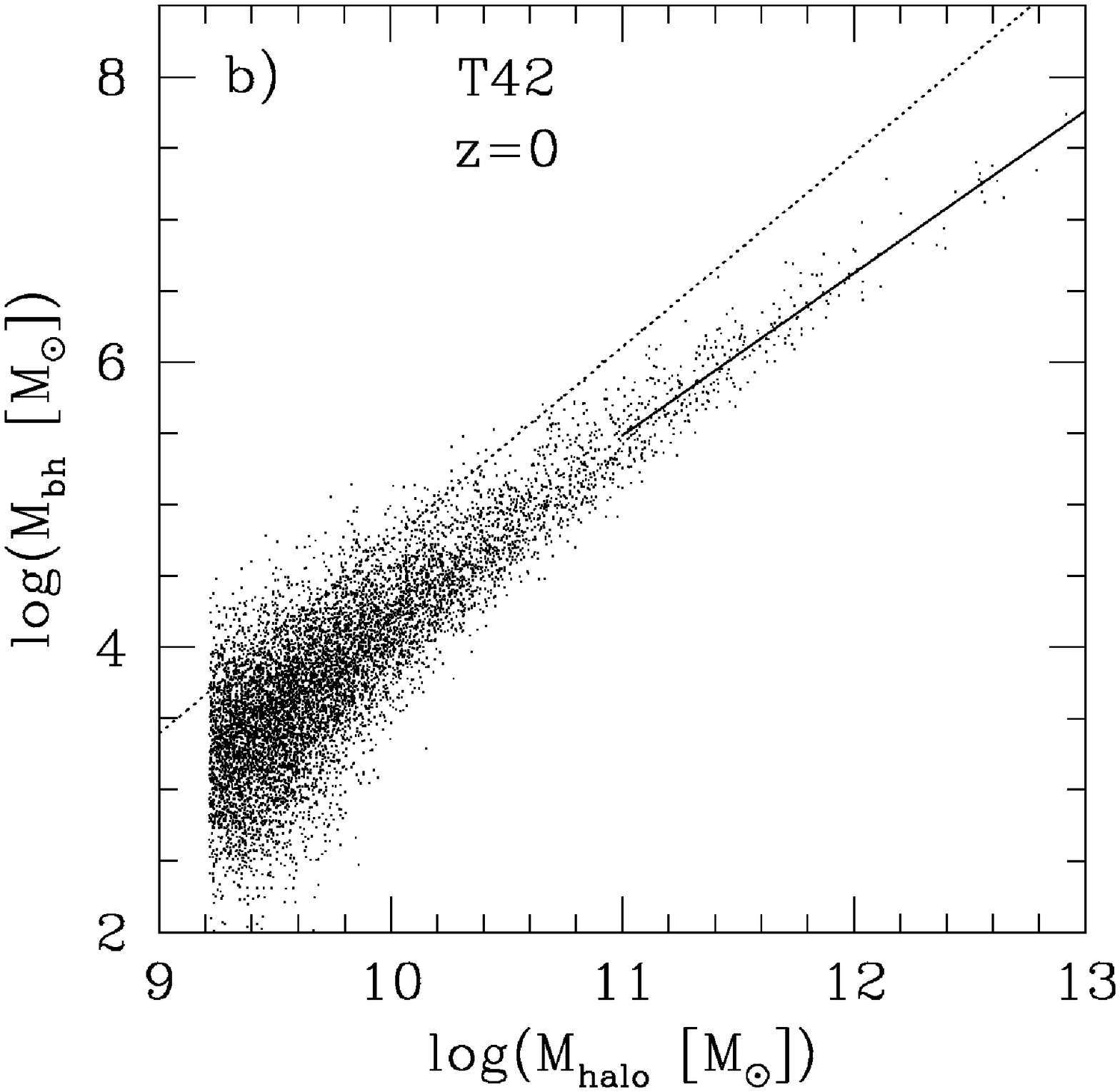,height=2.3in}
\end{displaymath}
\end{minipage}
\end{center}
\caption{\label{fig:three} $M_{\rm bh}$--$M_{\rm halo}$ end
distributions in (a) a model with maximal losses to gravitational
waves (TadK) and (b) a model with losses to gravitational waves
appropriate for a quasar population with fast--spinning black holes
(T42).  }
\end{figure}

Figure~\ref{fig:three} shows the additional effects that gravitational
wave losses have on the $M_{\rm bh}$--$M_{\rm halo}$ distribution.  In
Fig.~\ref{fig:three}a, the most extreme (and arguably unrealistic)
prescription for gravitational wave losses was adopted, with dramatic
consequences for the $M_{\rm bh}$--$M_{\rm halo}$ relation, especially
at large masses.  A least square fit is no longer satisfactory to
represent the distribution and, more importantly, all the massive BHs
in the model have lost most of their mass. As a result, the value of
$\rho_{\rm BH}$ has dropped by more than one order of magnitude from
$z=3$ to $z=0$ (compare T-forced at $z=3$ with TadK in
Table~\ref{tab:one}). While grossly exaggerated, the TadK model is
useful to highlight the consequences of gravitational wave
losses. Similar conclusions hold for the TadS model, even if the
effects are not as pronounced (not shown here; see
Table~\ref{tab:one}).\footnote{We note that Yu \& Tremaine (2002) have
included strong losses to gravitational waves in their discussion of
the quasar mass budget, but their description of mergers was
parametrized.}

Figure~\ref{fig:three}b shows the $M_{\rm bh}$--$M_{\rm halo}$ end
distribution in model T42, which may be considered as more realistic
for a population of quasars with systematically very fast-spinning BHs
(see discussion in \S2.3). The total decrease in $\rho_{\rm BH}$ at
$z=0$ is $\sim 30 \%$ and the flattening of the $M_{\rm bh}$--$M_{\rm
halo}$ relation is more pronounced than without any gravitational wave
losses (compare T42 with T0 in Table~\ref{tab:one}). Again, these
effects clearly dominate for the largest mass BHs and halos.  As
before, this is simply because of the larger number of mergers
experienced by more massive halos. Since mass is lost to gravitational
waves in each merger, the BHs in these massive halos will accumulate a
more significant decrease in their final mass at $z=0$.

\begin{figure}[t]
\begin{center}
\hspace{-1.7cm}
\begin{minipage}[t]{0.53\hsize}
\begin{displaymath}
\psfig{figure=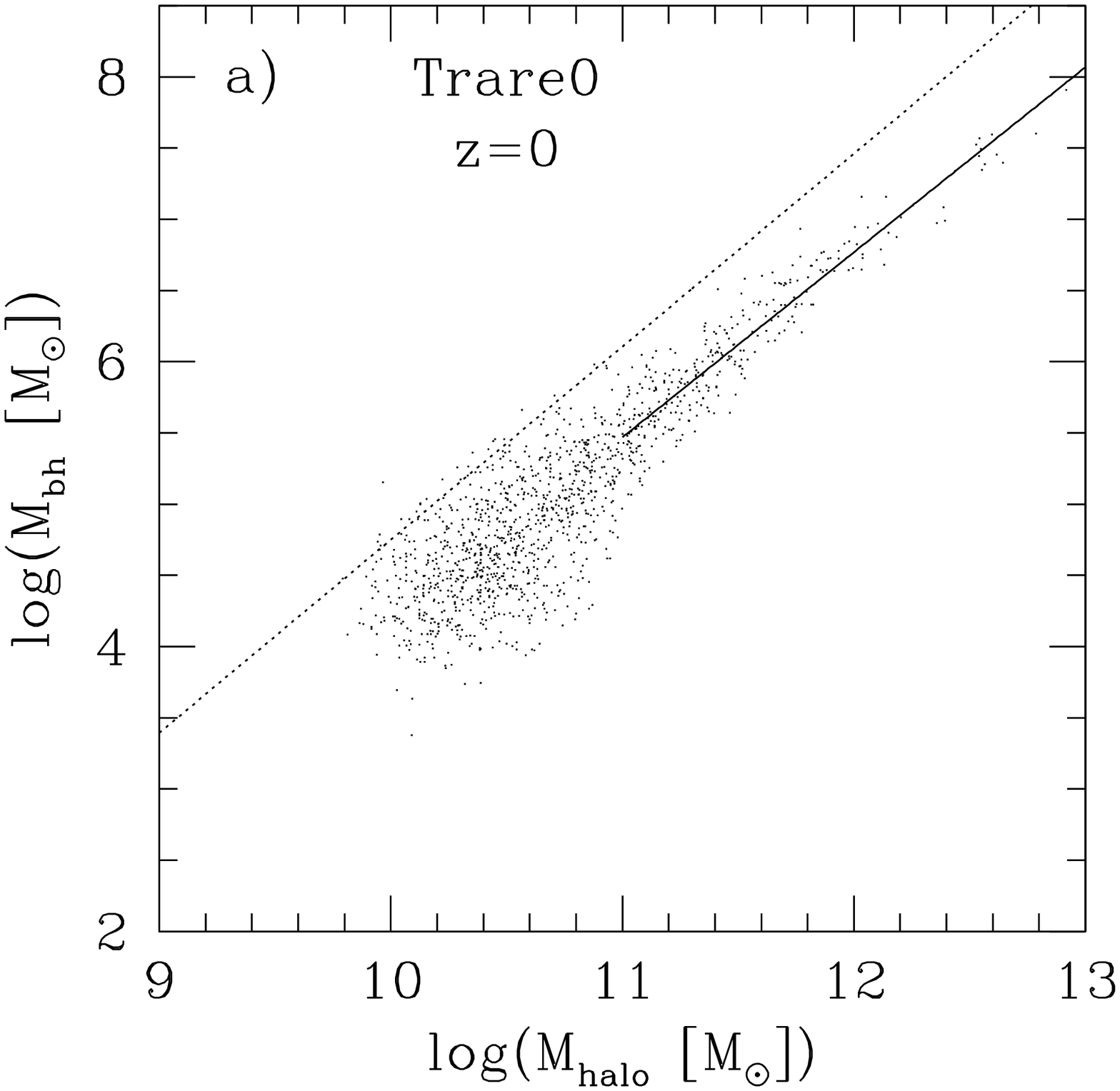,height=2.3in}
\end{displaymath}
\end{minipage}
\begin{minipage}[t]{0.60\hsize}
\begin{displaymath}
\psfig{figure=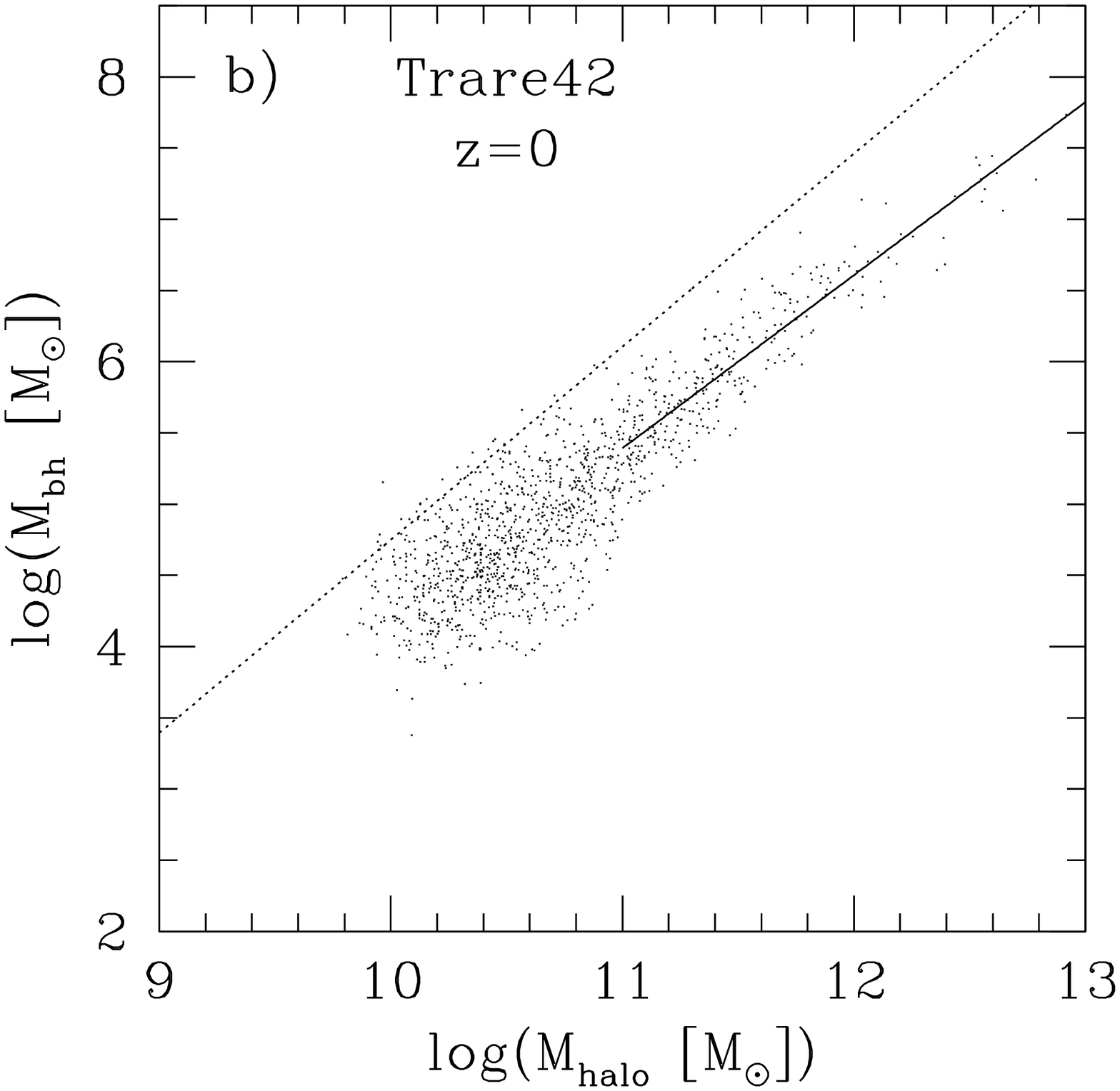,height=2.3in}
\end{displaymath}
\end{minipage}
\end{center}
\caption{\label{fig:four} $M_{\rm bh}$--$M_{\rm halo}$ end
distributions in (a) a model without any loss to gravitational waves
but a ten--times rarefied population of quasars, with black holes
initially located in the most massive halos at $z=3$ (Trare0) and (b)
the same model with losses to gravitational waves appropriate for a
quasar population with fast--spinning black holes (Trare42).}
\end{figure}

We have already noted that quasar populations could be rarer than
assumed in our standard models. Given indications that mergers
preferentially influence the properties of BHs at large masses, it is
important to verify whether our results still hold for a rarer
population of quasars. A specific class of ``rare'' models that we
have investigated are models in which the total number of BHs is
reduced by a factor $10$ and the remaining $10\%$ are forced to
systematically populate the $10\%$ most massive halos present in the
tree at $z=3$.\footnote{Rare BHs do tend to populate the most massive
halos after they experienced a large enough number of cosmological
mergers, as shown in the models of Menou et al. (2001).}  Because the
mass density in BHs is dominated by large masses for the mass
prescriptions we have adopted, the value of $\rho_{\rm BH}$ in these
rare models is only reduced by a small fraction as compared to models
with large quasar populations (compare T-forced with Trare-forced at
$z=3$ in Table~\ref{tab:one}). Note that we have also investigated a
second class of models in which an equally rare population of BHs ($10
\%$) populates, this time randomly, all the halos present in the tree
at $z=3$. In these models, the $M_{\rm bh}$--$M_{\rm halo}$ relation
initially in place at $z=3$ is rapidly wiped out by successive
cosmological mergers when the few galaxies hosting BHs experience
mergers with more massive, BH-free galaxies. We therefore consider
this second class of models with rare BHs as implausible.

Figure~\ref{fig:four} shows our results for $M_{\rm bh}$--$M_{\rm
halo}$ distributions in models with a rare population of quasars such
that BHs are initially located in the most massive halos at
$z=3$. Model Trare0 (Fig.~\ref{fig:four}a; no loss to gravitational
waves) shows that one consequence of such a rare population of quasars
is a noticeable reduction in the flattening effect of cosmological
mergers on the $M_{\rm bh}$--$M_{\rm halo}$ relation (see also
Table~\ref{tab:one}). Still, model Trare42 (Fig.~\ref{fig:four}b)
shows that when the effects of gravitational wave losses are included,
significant flattening of the $M_{\rm bh}$--$M_{\rm halo}$ relation
persists. The decrease in $\rho_{\rm BH}$ due to gravitational wave
losses also remains significant in models with a rare population of
quasars, $\sim 27 \%$ (compare Trare-forced at $z=3$ with Trare42 in
Table~\ref{tab:one}), even if the effect is less severe because of a
reduction in the total number of BH mergers in these models.

\subsection{Additional Results}

We have investigated a few other models in which a BH was assumed to
be present in each halo; the results from these models are listed in
Table~\ref{tab:one}. Model T6, which incorporates smaller losses to
gravitational waves during BH coalescences, still shows noticeable
flattening of the $M_{\rm bh}$--$M_{\rm halo}$ relation and some
reduction in the value of $\rho_{\rm BH}$. Models Tq3 and Tq2, on the
other hand, show that the effects of inefficient dynamical friction,
which preferentially affect small mass ratios and thus small mass BHs,
are not very important (compare Tq3 and Tq2 with T0 in
Table~\ref{tab:one}).

\begin{figure}[t]
\begin{center}
\hspace{-1.7cm}
\begin{minipage}[t]{0.53\hsize}
\begin{displaymath}
\psfig{figure=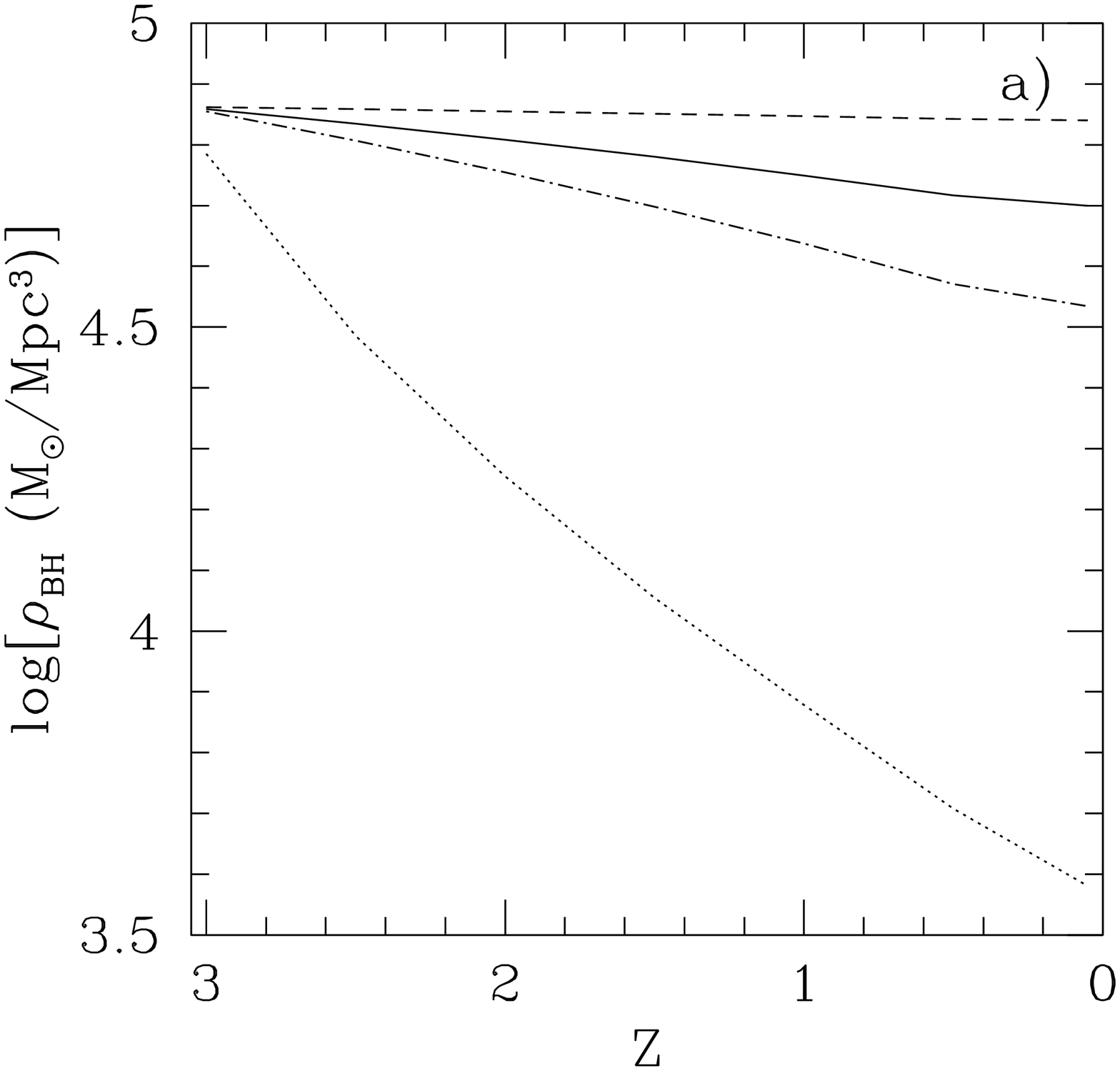,height=2.3in}
\end{displaymath}
\end{minipage}
\begin{minipage}[t]{0.60\hsize}
\begin{displaymath}
\psfig{figure=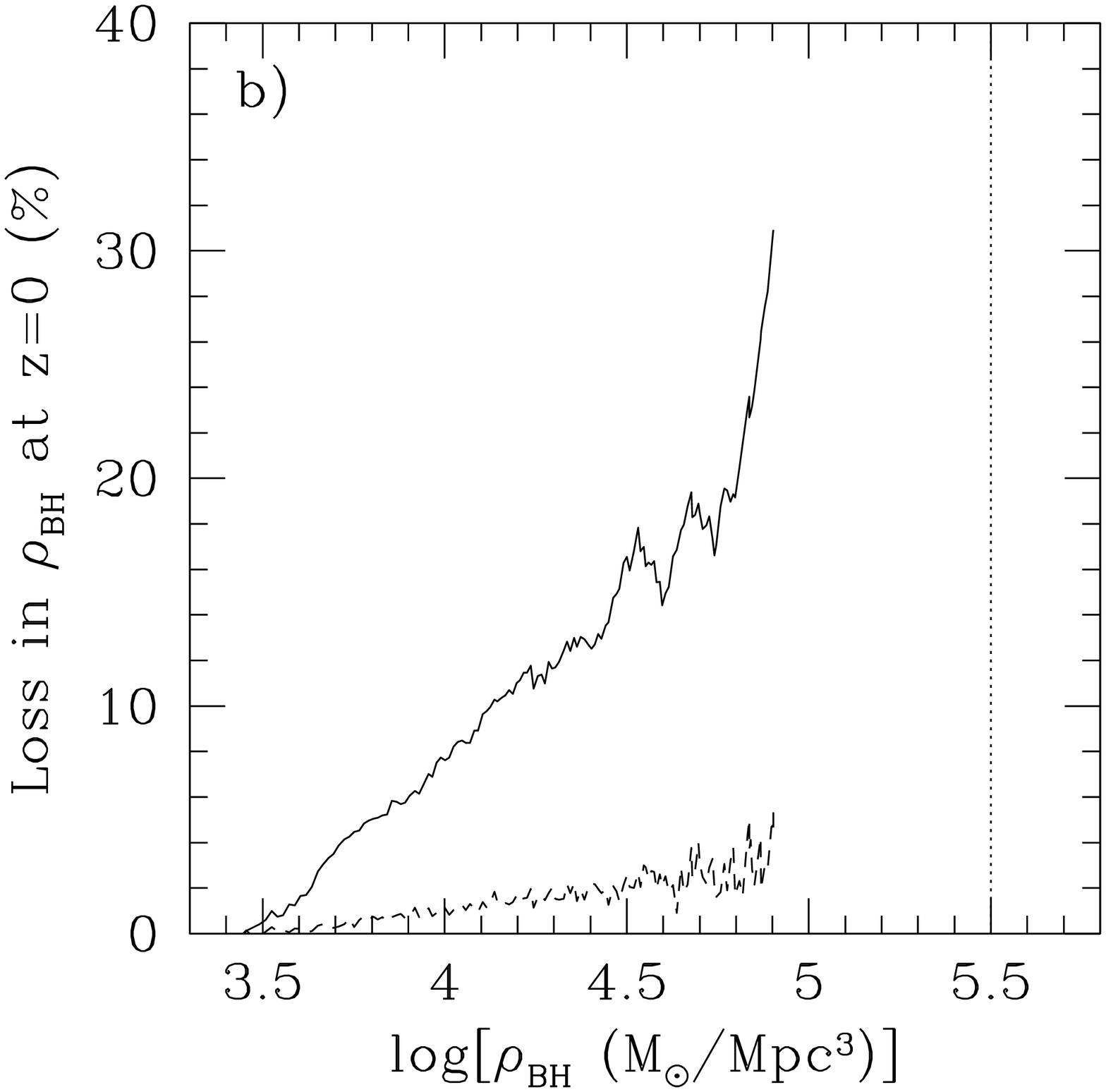,height=2.3in}
\end{displaymath}
\end{minipage}
\end{center}
\caption{\label{fig:five} (a) Evolution via successive mergers of the
black hole comoving mass density, $\rho_{\rm BH}$, as a function of
redshift, according to different prescriptions for gravitational wave
losses (model T6: dashed line; model T42: solid; model TadS:
dash-dotted; model TadK: dotted). (b) Percentage of the black hole
mass density lost at $z=0$, as a function of the mass density itself
(model T6: dashed; model T42: solid). An extrapolation of our results
to the currently-measured value, $\rho_{\rm BH} \sim 3 \times
10^5$~M$_\odot$~Mpc$^{-3}$ (vertical dotted line), suggests that
losses $> 30 \%$ are possible.}
\end{figure}

While we have found some evidence that the magnitude of flattening of
the $M_{\rm bh}$--$M_{\rm halo}$ relation due solely to cosmological
mergers may be reduced for a rare and massive population of quasars,
the effects due to gravitational wave losses appear to be important in
all cases, especially in terms of the $\rho_{\rm BH}$
deficit. Figure~\ref{fig:five}a shows, for models T6, T42, TadS and
TadK, that this deficit is related to mergers which occur over a wide
range of redshifts and therefore does not strongly depend on our
specific choice of $z=3$ as the initial redshift in our models (even
if starting at a smaller initial redshift would obviously reduce the
value of the $\rho_{\rm BH}$ deficit at $z=0$).

We have seen previously that the $\rho_{\rm BH}$ deficit due to
gravitational wave losses is preferentially due to losses at large BH
masses (see, e.g., Fig.~\ref{fig:three}a), because these BHs reside in
larger mass halos which experience a larger number of mergers, on
average. We have also already commented on the limiting size of the
merger tree used, which results in a relative scarcity of very massive
halos and is the reason behind the systematically small values of
$\rho_{\rm BH}$ found in our models, as compared to the
observationally inferred value. It is natural to wonder, then, if
values for the $\rho_{\rm BH}$ deficit due to gravitational wave
losses could also be underestimated in our models because of the
statistical scarcity of massive halos.

We have attempted to answer this question by measuring how the
$\rho_{\rm BH}$ deficit at $z=0$ depends on the range of halo masses
included in the merger tree. Figure~\ref{fig:five}b shows graphically
the result of this exercise and confirms our expectations. The total
$\rho_{\rm BH}$ deficit accumulated at $z=0$ (expressed in $\%$ loss
relative to the no-loss value) is shown for models T6 and T42, as a
function of the value taken by $\rho_{\rm BH}$ as we remove an
increasingly large number of the most massive halos present in the
tree at $z=0$. Although this test cannot replace a full investigation
with a larger tree, it does show that most of the $\rho_{\rm BH}$
deficit is related to the few most massive BHs and halos and it
suggests, via an extrapolation to $\rho_{\rm BH} \sim 3 \times
10^5$~M$_\odot$~Mpc$^{-3}$ (the measured value, indicated by a
vertical solid line in Fig.~\ref{fig:five}b) that $\rho_{\rm BH}$
deficits $> 30 \%$ are expected in T42 models constructed with larger
trees.

%%%%%%%%%%%%%%%%%%%%%%%%%%%%%%%%%%%%%%%%%%%%%%%%%%%%%%%%%%%%%
%                                                           %
% You may repeat \section{SECTION N-th HEADING TYPE HERE}   %
%                                                           %
% Do start a subsection or sub-subsection, do this:-        %
%                                                           %
%   \subsection{SUBSECTION HEADING TYPE HERE}               %
%                                                           %
%   \subsubsection{SUBSUBSECTION HEADING TYPE HERE}         %
%                                                           %
% instead of the above                                      %
%                                                           %
%%%%%%%%%%%%%%%%%%%%%%%%%%%%%%%%%%%%%%%%%%%%%%%%%%%%%%%%%%%%%

\section{Discussion and Conclusion}

In this study, we have isolated and quantified the effects that
repeated galactic mergers and BH binary coalescences may have on a
population of quasars in a cosmological context. While the
characteristics of the local population of dead quasars is relatively
well known, it is not the case for more distant quasars and we have
therefore represented these quasars with a variety of plausible
populations of massive BHs (T- and FWL-models, Trare models in
Table~\ref{tab:one}).

Our models indicate that galactic mergers alone (excluding
gravitational wave losses) influence somewhat the properties of the
quasar population, by redistributing BHs in galaxies and flattening
the $M_{\rm bh}$--$M_{\rm halo}$ relation. The effect appears to be
small for a rare population of BHs preferentially located in massive
galaxies (as may be expected, for instance, from recoil effects;
Favata et al. 2004). This lends support to the idea that the
high-redshift quasar population has properties which are rather
similar to those of local dead quasars.

According to our models, however, losses to gravitational waves during
repeated BH binary coalescences have the potential to modify
substantially the characteristics of the quasar population. First,
they contribute to the flattening of the $M_{\rm bh}$--$M_{\rm halo}$
relation by preferentially reducing the mass of the largest BHs, since
these BHs experience a larger number of mergers. Second, losses to
gravitational waves systematically reduce the BH mass density,
$\rho_{\rm BH}$, over cosmic times. This potentially important
cumulative effect reaches up to $30 \%$ of the no--loss value of
$\rho_{\rm BH}$ in our models with a population of maximally rotating
BHs, and we have argued that the effect could be even stronger had we
used a larger cosmological merger tree.

One must keep in mind that our models are idealized in several
ways. We have neglected the growth in BH mass due to accretion. We
note, for instance, that in models describing accretion, a flattening
of the $M_{\rm bh}$--$M_{\rm halo}$ relation could be counter-balanced
by prescribing accretion onto massive BHs in such a way as to
reproduce the apparently similar relation existing between BHs and
their host galaxies at $z=0$ and at $z=3$ (Shields et al. 2003). Our
estimation of the effects due to gravitational wave losses is also
very much simplified by adopting idealized loss prescriptions and
assuming that all BH binaries coalesce efficiently with a same
prescribed loss.

Still, our models suggest that the issue of losses to gravitational
waves may be important for the interpretation of the quasar mass
budget. A deficit in the mass budget of local dead quasars relative to
that of distant active quasars has usually been interpreted as
resulting from highly efficient BH accretion during active quasar
phases ($\epsilon > 0.1$; see \S1). The results presented here
indicate that part of this local mass deficit could be attributed
instead to gravitational wave losses accumulated over cosmic times
during repeated BH binary coalescences. In a companion study (Menou \&
Haiman 2004), we put more stringent limits on the role of
gravitational wave losses in modifying the mass budget of merging
quasars, by using more accurate loss prescriptions based on detailed
general relativistic calculations.

%%%%%%%%%%%%%%%%%%%%%%%%%%%%%%%%%%%%%%%%%%%%%%%%%%%%%%%%%%%%%
% Doing Acknowledgement                                     %
%%%%%%%%%%%%%%%%%%%%%%%%%%%%%%%%%%%%%%%%%%%%%%%%%%%%%%%%%%%%%

\section*{Acknowledgments}

K.M. thanks the Department of Astronomy at the University of Virginia
for their hospitality. Z.H. was supported in part by NSF through
grants AST-0307200 and AST-0307291.

%%%%%%%%%%%%%%%%%%%%%%%%%%%%%%%%%%%%%%%%%%%%%%%%%%%%%%%%%%%%%
% Doing Appendix(ices)                                      %
%%%%%%%%%%%%%%%%%%%%%%%%%%%%%%%%%%%%%%%%%%%%%%%%%%%%%%%%%%%%%

%\appendix

%\section{HEADING FOR APPENDIX A}

%\renewcommand{\theequation}{A.\arabic{equation}}

%TYPE TEXT FOR APPENDIX A HERE.

%\section{HEADING FOR APPENDIX B}

%\renewcommand{\theequation}{B.\arabic{equation}}

%TYPE TEXT FOR APPENDIX B HERE.

%%%%%%%%%%%%%%%%%%%%%%%%%%%%%%%%%%%%%%%%%%%%%%%%%%%%%%%%%%%%%
% Doing references:                                         %
%%%%%%%%%%%%%%%%%%%%%%%%%%%%%%%%%%%%%%%%%%%%%%%%%%%%%%%%%%%%%

\end{document}